\documentstyle[pra,aps,twocolumn,psfig,floats]{revtex}
\draft
\begin{document}

\title{Conditional generation of sub-Poissonian light from two-mode
squeezed vacuum \\ via balanced homodyne detection on idler mode}

\author{Jarom\'{\i}r Fiur\'{a}\v{s}ek}

\address{Department of Optics, Palack\'{y} University, 17. listopadu 50,
77200 Olomouc, Czech Republic}

\maketitle

\begin{abstract}
A simple scheme for conditional generation of nonclassical light with
sub-Poissonian photon-number statistics is proposed.
The method utilizes entanglement of signal and idler modes in
two-mode squeezed vacuum state generated in optical parametric
amplifier. A quadrature component of the idler mode is measured in
balanced homodyne detector and only those experimental runs where
the absolute value of the measured quadrature is higher than certain
threshold are accepted. If the threshold is large enough then the
conditional output state of signal mode exhibits reduction of
photon-number fluctuations below the coherent-state level.
\end{abstract}


\section{Introduction}

The possibility of a {\em conditional generation}
of nonclassical states of light has been intensively investigated
in recent years. The conditional quantum-state preparation schemes
benefit from quantum
entanglement between signal mode and an ancilla system (e.g. two output
ports of a beam-splitter \cite{Ban96}, signal and idler modes in parametric
down-conversion \cite{Watanabe88,Luis98},
or cavity mode and an atom in cavity QED \cite{Vogel93}).
Suppose we measure some observable of the ancilla. The collapse of the
ancilla state, caused by the measurement, influences, due to
entanglement, the state of the signal mode.
In the conditional quantum-state preparation schemes we accept only those
experimental runs where the required measurement outcome (or, more
generally, a sequence of measurement outcomes) is observed, and we
reject all unsuccessful trials.

Based on this general strategy, schemes for conditional generation of
Fock states \cite{DAriano00}, Schr\"{o}dinger cat states \cite{Song90,Dakna97}
and arbitrary superpositions of Fock states in a cavity mode
\cite{Vogel93,Harel96} have been suggested. In particular, photon-number
detection on idler mode of two-mode squeezed vacuum allows one to conditionally
generate a highly nonclassical state of signal mode whose Wigner
function may be negative near the origin of the phase space
\cite{Lvovsky01,Paris01}. Alternatively, one can also use the detected
intensity of idler beam as a negative feedback which modulates the intensity
of pumping beam \cite{Tapster88}, or as a trigger of an optical shutter placed
in path of pump  or signal beams \cite{Walker85}. All these techniques
can provide a sub-Poissonian light in signal beam.

A direct detection of the number of photons of the idler mode is not the only
possibility here. A quadrature-measurement via balanced homodyne
detection would prepare the signal in squeezed coherent state
\cite{Watanabe88,Luis98}. Similarly, a heterodyne detection on idler
leads to coherent state of the signal \cite{Watanabe88}.
Unfortunately, the coherent amplitudes are random, because they are
proportional to the value of the detected quadrature of the idler mode.
This problem may be circumvented by the feedforward technique
where a suitable displacement of the signal is performed thereby
compensating the undesired effects of the measurement on the
idler mode \cite{Watanabe88}.

In this paper we propose an alternative simple and new way of conditional
generation of nonclassical light in the signal mode via homodyne detection
of the idler mode. In the suggested scheme, the experimental run is
accepted only when the measured quadrature falls within a certain chosen
window of quadratures. This guarantees that the
conditional state is prepared with finite probability, controlled by the
size and position of the quadrature window. For the sake of concretness,
we shall assume that the signal state is successfully generated whenever
the absolute value of the detected idler quadrature is higher than certain
threshold $x_0$. We shall see that the conditional state of signal can
exhibit sub-Poissonian photon-number statistics. The amount of suppression of
photon-number fluctuations can be controlled by $x_0$.
There is a trade-off between preparation-probability and  the
amount of photon-number fluctuations of the generated state.
Strong suppression of the fluctuations can be achieved at the expense of
a small success rate and vice versa.
We address the influence of imperfect detection and show that the
nonclassical light can be prepared only if the total efficiency of
homodyning is higher than certain threshold.

The paper is organized as follows. In Sec. II we provide a description
of the suggested experimental setup. The quantum statistical properties
of the conditional state of signal mode are studied in Sec. III.
The influence of imperfect homodyne detection is discussed in Sec. IV.
Finally, Sec. V contains conclusions.

\section{Conditional generation}

Consider the experimental setup shown in Fig. 1.
The two-mode squeezed vacuum state $|\psi\rangle$
prepared in the nondegenerate parametric amplifier (NOPA) reads
\cite{Schumaker85}
\begin{eqnarray}
|\psi\rangle &\equiv& \exp\left(r\hat{a}_s^\dagger
\hat{a}_i^\dagger-r\hat{a}_s \hat{a}_i\right) |0 \rangle_s |0 \rangle_i
\nonumber \\
&=&\sqrt{1-\lambda}
\sum_{n=0}^\infty  \lambda^{n/2} |n\rangle_s|n\rangle_i,
\label{twomodevac}
\end{eqnarray}
where $\hat{a}_s$, $\hat{a}_{i}$ ($\hat{a}_s^\dagger$,
$\hat{a}_i^\dagger$) are annihilation (creation) operators  of signal
and idler modes, respectively, $|n\rangle_s$, $|n\rangle_i$  are Fock states
of the signal and idler modes, $r$ is (real) squeezing constant of NOPA
and $\lambda=\tanh^2 r$. We note that the currently available
parametric amplifiers can produce strongly squeezed light.
For example, Ayt\"{u}r and Kumar \cite{Kumar90}
reported a parametric gain $\cosh^2 r \approx 10$, which corresponds
to $\lambda\approx 0.9$.

As illustrated in Fig. 1, the idler mode of two-mode squeezed vacuum is fed
to the balanced homodyne detector. The measurement on idler triggers
the observation of the signal mode, which is performed only if the detected
idler quadrature falls within the chosen window of quadratures $\cal{X}$.
In this way we conditionally generate the output state of the signal
mode.

Since it is experimentally challenging to maintain
a fixed phase of the local oscillator, we consider the
homodyne detection with randomized phases.
Positive operator-valued measure (POVM)
describing this type of quantum measurements reads
\begin{equation}
\hat{\Pi}_i(x)=\frac{1}{2\pi}\int_{0}^{2\pi}
|x;\phi\rangle_i\langle x;\phi| d \phi,
\end{equation}
where $|x;\phi\rangle_i$  is an eigenstate of the quadrature operator
$\hat{x}_{i,\phi}=(\hat{a}_i e^{-i\phi}+\hat{a}_i^\dagger e^{i\phi})/\sqrt{2}$,
\begin{equation}
\hat{x}_{i,\phi} |x;\phi\rangle_i = x|x;\phi\rangle_i.
\end{equation}
The POVM corresponding to the quadrature window $\cal{X}$
can be expressed as
\begin{equation}
\hat{\Pi}_i=\int_{\cal{X}}\hat{\Pi}_i(x) dx.
\label{povmi}
\end{equation}
The (normalized) reduced density matrix $\hat{\rho}_s$ of the signal
can be obtained by tracing over idler mode
\begin{equation}
\hat{\rho_s}= \frac {{\rm Tr}_i\left(\hat{ I}_s \otimes \hat{\Pi}_i
|\psi\rangle\langle \psi|\right)}{
\langle \psi|\hat{ I}_s \otimes\hat{\Pi}_i|\psi\rangle},
\label{rhoout}
\end{equation}
where $\hat{I}_s$ is  unit operator on Hilbert space of signal mode.

Due to the phase randomized homodyning,
density matrix of the conditional state of the signal mode
is diagonal in Fock basis,
\begin{equation}
\hat{\rho}_s=\sum_{n=0}^\infty p_n |n\rangle_s\langle n|.
\end{equation}
On inserting the POVM (\ref{povmi}) into Eq. (\ref{rhoout}),
we get  formula for the photon-number distribution $p_n$,
\begin{equation}
p_n= \frac{1-\lambda}{C(\lambda,{\cal{X}})} \lambda^{n} q_n,
\label{pndistribution}
\end{equation}
where
\begin{equation}
C(\lambda,{\cal{X}})\equiv
\langle \psi|\hat{ I}_s \otimes\hat{\Pi}_i|\psi\rangle
=(1-\lambda)\sum_{n=0}^\infty \lambda^n q_n
\label{cdef}
\end{equation}
is a normalization coefficient and
\begin{equation}
q_n =\int_{0}^{2\pi}\frac{d\phi}{2\pi}\int_{\cal{X}}
|\langle x;\phi|n\rangle|^2 dx
=\int_{\cal{X}}\frac{H_n^2 (x) e^{-x^2}}{2^n n! \sqrt{\pi}}  dx.
\end{equation}
Here $H_n(x)$ denotes Hermite polynomial of variable $x$.

\begin{figure}[!t!]
\centerline{\psfig{figure=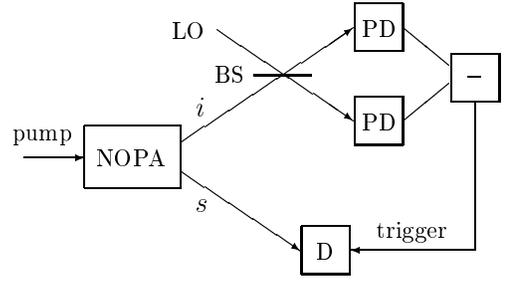,width=0.76\linewidth}}

\vspace*{2mm}

\caption{Scheme of con\-ditio\-nal gene\-ration of sub-Poisso\-nian light
in signal mode from two-mode squeezed vacuum prepared in nondegenerate
parametric amplifier (NOPA). A quadrature of idler mode is measured in
balanced homodyne detector, which consists of a beam splitter BS,
strong coherent local oscillator LO, two photodetectors PD and a
subtracter of the two signals. Detected quadrature falling
within the chosen window $\cal{X}$
triggers the observation of the signal mode.}
\end{figure}

\section{Photon-number statistics of the conditional state}

So far we have assumed an arbitrary  $\cal{X}$. In what follows, we
analyze the case where the conditional state of signal mode is
successfully generated if the absolute value of the  idler quadrature
is higher than certain lower bound $x_0$. In this case we can write,
\begin{equation}
q_n=\frac{2}{2^n n! \sqrt{\pi}}\int_{x_0}^\infty H_n^2 (x) e^{-x^2} dx.
\label{qn}
\end{equation}
The normalization factor $C(\lambda,x_0)$ represents the probability of
conditional generation of the output state (\ref{rhoout}),
i.e. the probability
that the absolute value of the measured quadrature of the idler would
be larger than $x_0$. The idler mode is in thermal (chaotic) state and
its quadrature component $\hat{x}_{i,\phi}$ exhibits Gaussian distribution
with zero mean and variance $(1+\lambda)/[2(1-\lambda)]$.
The probability that $|x|>x_0$ can be thus expressed as
\begin{equation}
C(\lambda,x_0)= \frac{2}{\sqrt{\pi}}
\sqrt{\frac{1-\lambda}{1+\lambda}}
\int_{x_0}^\infty \exp\left(-x^2\frac{1-\lambda}{1+\lambda}\right) dx.
\end{equation}
After a straightforward integration we arrive at
\begin{equation}
C(\lambda,x_0)=
1-{\rm erf} \left(x_0\sqrt{\frac{1-\lambda}{1+\lambda}}\right),
\label{c}
\end{equation}
where the error function is defined as
\begin{equation}
{\rm erf}(x)=  \frac{2}{\sqrt{\pi}}\int_0^x e^{-y^2} dy.
\end{equation}
The function $C(\lambda,x_0)$ contains a complete information about the
quantum-statistical properties of the state (\ref{rhoout}). From the
definition (\ref{cdef}) we can deduce that  $C(\lambda,x_0)$ is a generating
function of moments of the photon number distribution
(\ref{pndistribution}),
\begin{equation}
\langle \hat{n}^k\rangle=
\frac{1-\lambda}{C(\lambda,x_0)}
\left(\lambda\frac{d}{d\lambda}\right)^k
\left[\frac{C(\lambda,x_0)}{1-\lambda}\right].
\label{nmom}
\end{equation}
Also the normally ordered moments can be determined from $C(\lambda,x_0)$,
\begin{equation}
\langle:\hat{n}^k:\rangle \equiv
\left\langle \frac{\hat{n}!}{(\hat{n}-k)!}\right\rangle=
\frac{(1-\lambda)\lambda^k}{C(\lambda,x_0)}  \frac{d^k}{d \lambda^k}
\left[\frac{C(\lambda,x_0)}{1-\lambda}\right].
\label{normalmom}
\end{equation}
In particular, we obtain expressions for the mean photon number
\begin{equation}
\langle \hat{n} \rangle =
\frac{\lambda}{1-\lambda}
+\frac{2\lambda x_0C^{-1}(\lambda,x_0)}{\sqrt{\pi}
[(1-\lambda)(1+\lambda)^3]^{1/2}}
\exp\left(-x_0^2\frac{1-\lambda}{1+\lambda}\right)
\label{nmean}
\end{equation}
and second factorial moment
\begin{eqnarray}
\langle :\! \hat{n}^2 \! :\rangle&=&
\frac{2 \lambda^2}{(1-\lambda)^2} +
\sqrt{\frac{1-\lambda}{1+\lambda}}
\frac{2 x_0 \lambda^2 }{\sqrt{\pi}(1-\lambda^2) C(\lambda,x_0)}
\nonumber \\[1.5mm] && \times
\left[\frac{1+4\lambda}{1-\lambda^2}+\frac{2
x_0^2}{(1+\lambda)^2}\right]
\exp\left(-x_0^2\frac{1-\lambda}{1+\lambda}\right).
\label{sfmoment}
\end{eqnarray}

\begin{figure}[!t!]
\centerline{\psfig{figure=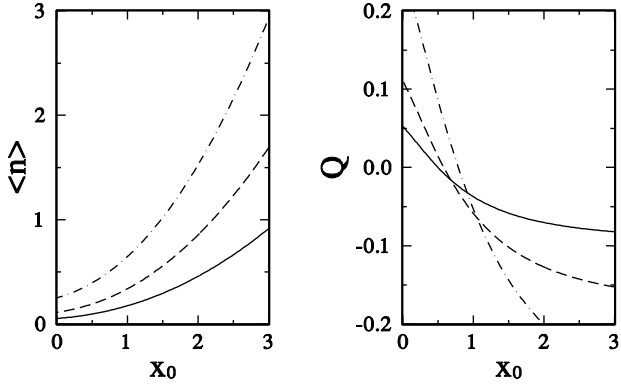,width=0.95\linewidth}}

\vspace*{2mm}

\caption{Mean photon number $\langle \hat{n} \rangle$
and Mandel $Q$-factor versus $x_0$  for three different squeezing
parameters $\lambda=0.05$ (solid line), $\lambda=0.1$ (dashed
line) and $\lambda=0.2$ (dot-dashed line).}
\end{figure}

The conditional output states $\hat{\rho}_s$ can be highly nonclassical,
exhibiting sub-Poissonian statistics. The suppression of photon-number
fluctuations can be conveniently characterized by Mandel $Q$-factor
\cite{Mandel95}
\begin{equation}
Q=\frac{\langle :\! (\Delta \hat{n})^2 \! : \rangle}{\langle \hat{n}\rangle}=
\frac{\langle \hat{n}^2\rangle -\langle \hat{n}\rangle^2}
{\langle \hat{n}\rangle}-1.
\label{q}
\end{equation}
It holds that $Q \geq -1$ and the equality is achieved for Fock
state. The light is sub-Poissonian when the photon-number variance
$\langle (\Delta \hat{n})^2 \rangle $ is less than the mean number of
photons $\langle \hat{n}\rangle$. This is indicated by negative value of
$Q$. The statistics are Poissonian when $Q=0$ and super-Poissonian if
$Q>0$. 
\begin{figure}[!t!]
\centerline{\psfig{figure=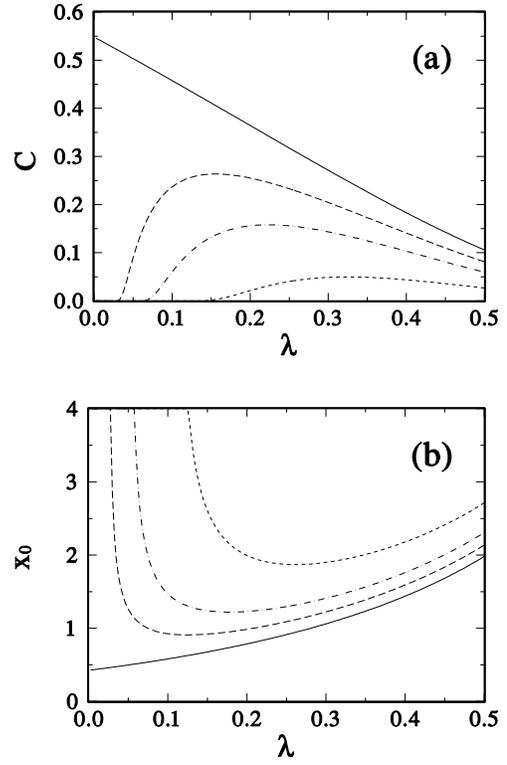,width=0.75\linewidth}}

\vspace*{2mm}

\caption{(a) The probability of conditional generation of state with
$Q=0$ (solid line), $Q=-0.05$ (dashed line) $Q=-0.1$
(dot-dashed line), and $Q=-0.2$ (short-dashed line).
Figure (b) shows the corresponding values of $x_0$.}
\end{figure}

The dependence of the mean photon number (\ref{nmean}) and the $Q$-factor
(\ref{q}) on the lower bound $x_0$ is plotted in Fig. 2
for three different values of $\lambda$.
We can see that $\langle \hat{n}\rangle$ monotonically
grows with $x_0$ while $Q$ monotonically decreases and eventually
becomes negative. The figure clearly indicates that the
states $\hat{\rho}_s$ can possess sub-Poissonian statistics. However,
we should be concerned with  the probability $C(\lambda,x_0)$
of generation of such nonclassical states.

Figure 3a shows dependence of this probability
on $\lambda$  for four different  $Q$-factors. The corresponding values
of $x_0$ required for reaching given $Q$ are displayed in Fig. 3b.
Consider first the boundary case when the photon-number
fluctuations are suppressed to the level of Poissonian distribution,
hence $Q=0$.  The probability of generation can be almost $55\%$ for weakly
squeezed states ($\lambda \ll 1$) and decreases with growing $\lambda$.
In agreement with Fig. 2b, the $x_0$ required for achieving
$Q=0$ increases with $\lambda$. It is interesting
that there is a lower bound on $x_0$ as $\lambda \rightarrow 0$.
In order to suppress the fluctuations to Poissonian level, we need
finite $x_0$ even for very weak squeezing. This minimum
$x_{0,\rm min}$ can be determined if we consider the formulas
(\ref{nmean}), (\ref{sfmoment}) and (\ref{q})
in the limit $\lambda \rightarrow 0$.
It follows that $x_{0,\rm min}$ is a root of a transcendent equation
\[
\left(1+\frac{2 x_0e^{-x_0^2}}{\sqrt{\pi}[1-{\rm erf}(x_0)]}\right)^2
=2+\frac{2x_0(1+2x_0^2)}{\sqrt{\pi}[1-{\rm erf}(x_0)]}e^{-x_0^2},
\]
whose numerical solution provides $x_{0,\rm min}=0.4248$.

Assume now that we  want to generate a nonclassical
state with negative $Q$. Figure 3 reveals that for each $Q<0$
there is certain optimal value of $\lambda$ which allows us to generate
this state with highest probability. The peak in probability roughly
corresponds to the minimum of $x_0$ but those two extrema do not
coincide.

\begin{figure}[!t!]
\centerline{\psfig{figure=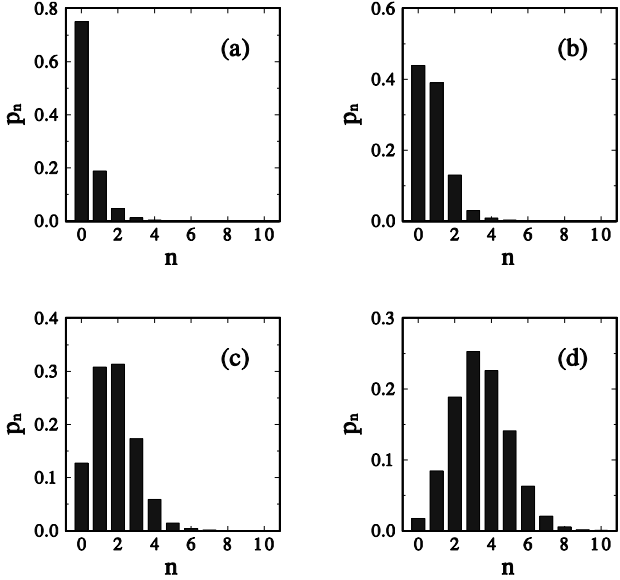,width=0.95\linewidth}}

\vspace*{2mm}

\caption{Photon-number distribution of the conditional state
$\hat{\rho}_s$ for $\lambda=0.25$ and (a) $x_0=0$, (b) $x_0=1$,
(c) $x_0=2$, and (d) $x_0=3$. }

\end{figure}

Having investigated the photon-number variance we turn our attention to
the photon-number distribution itself. The coefficients $q_n$ can
be calculated from $C(\lambda,x_0)$ as follows,
\begin{equation}
q_n= \left.\frac{1}{n!} \frac{d^n}{d\lambda^n}\left[
\frac{C(\lambda,x_0)}{1-\lambda}\right]\right|_{\lambda=0} .
\end{equation}
This formula, however, is not very convenient for practical
calculations. Therefore we present a different derivation of $q_n$
which leads to formulas suitable for numerical processing.
In order to perform the integration in Eq. (\ref{qn}), we invoke the
generating function for Hermite polynomials,
\begin{equation}
e^{-h^2+2hx}=\sum_{n=0}^\infty \frac{1}{n!} H_n(x) h^n
\end{equation}
and rewrite Eq. (\ref{qn}) as
\begin{equation}
q_n= \frac{1}{2^n n! }
\left. \frac{\partial^{2n}}{\partial h^n \partial k^n}
\int_{x_0}^{\infty} e^{-h^2+2hx} e^{-k^2+2kx}e^{-x^2} dx
\right|_{h=k=0}.
\end{equation}
After some algebra we find that
\begin{equation}
q_n= \frac{1}{2^n n! }
\left. \frac{\partial^{2n}}{\partial h^n \partial k^n}
\left[e^{2hk}\left[1-{\rm erf}\,(x_0-h-k)\right]\right]
\right|_{h=k=0}.
\end{equation}
On performing the differentiation we finally arrive at
\begin{equation}
q_n= 1-{\rm erf}(x_0) +
\frac{e^{-x_0^2}}{\sqrt{\pi}}
\sum_{j=0}^{n-1} \frac{2^{-j}}{(j+1)!} H_j(x_0)H_{j+1}(x_0).
\end{equation}
This expression looks particularly simply when
rewritten as a recurrence relation,
\begin{equation}
q_n=q_{n-1} +
\frac{e^{-x_0^2}}{\sqrt{\pi}2^{n-1} n!}
  H_{n-1}(x_0)H_{n}(x_0),
\end{equation}
starting with $q_0=1-{\rm erf}(x_0)$.

Figure 4 illustrates how
the shape of the photon number distribution of $\hat{\rho}_s$
changes with $x_0$. When $x_0=0$ and
all quadratures are accepted, the signal mode is in thermal state with
mean number of chaotic photons $\langle \hat{n}\rangle
=\lambda/(1-\lambda)$ and positive $Q$-factor ($Q_a=0.333$), see Fig. 4a.
This initially super-Poissonian distribution becomes narrower
when the lower bound of accepted quadratures $x_0$ is increased
and the statistics shown in Figs 4b-d are sub-Poissonian, as indicated
by negative $Q$-factors ($Q_b=-0.026$, $Q_c=-0.216$, and $Q_d=-0.297$).
 The selection  conditioned by quadrature measurements
suppresses the contribution of vacuum and low Fock states and  $p_n$
becomes peaked about $n$ which grows with $x_0$.

\begin{figure}[!t!]
\centerline{\psfig{figure=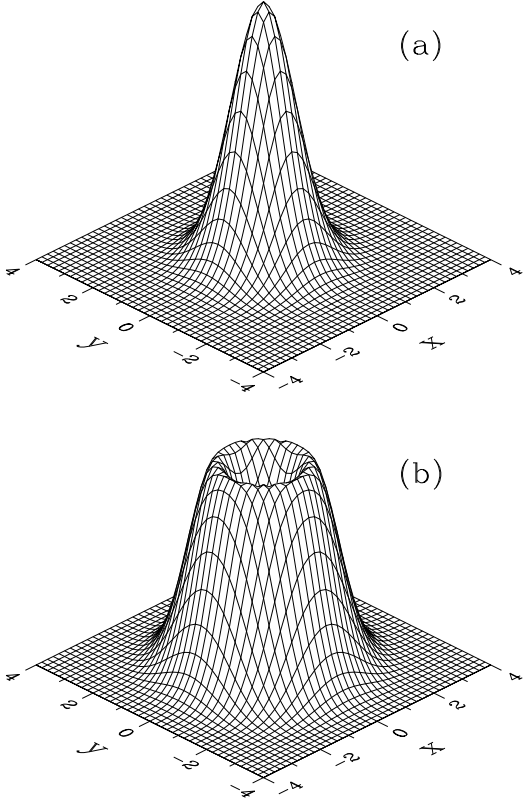,width=0.8\linewidth}}

\vspace*{2mm}
\caption{The Husimi quasidistributions of the state $\hat{\rho}_s$
for $\lambda=0.25$ and  (a) $x_0=0$, (b) $x_0=2$. }

\end{figure}

These changes in the form of the photon-number distribution are
reflected in the shape of quasidistributions, which are phase space
representations of the quantum state.
The quasidistributions for $\hat{\rho}_s$ are axially
symmetric around the origin of phase space because the density matrix
$\hat{\rho}_s$ is diagonal in Fock basis and does not carry any
information  about phase. The Husimi quasidistribution
$W_{\cal{A}}(\alpha)= \pi^{-1} \langle \alpha |\hat{\rho}_s
|\alpha\rangle$
corresponding to states from  Figs. 4a and 4c is plotted in Fig. 5.
The quasidistribution of thermal (chaotic) state shown in Fig. 5a
has Gaussian shape and is peaked at the origin of the phase space.
On the other hand, $W_{\cal{A}}$ depicted in Fig. 5b has a clear dip
in the origin and is peaked at a nonzero radius.

Although the conditional state $\hat{\rho}_s$ can be nonclassical,
its Wigner function $W_{s}(x_s,p_s)$ is always positive. This should be
contrasted with conditional generation based on photodetection on the idler,
which may lead to states with negative Wigner function
\cite{Lvovsky01,Paris01}. The positivity of $W_s$ can be proved
by recalling that the trace of two operators can be evaluated by
integrating the product of their Wigner functions over the whole phase
space. The Wigner representations of the two-mode squeezed vacuum
(\ref{twomodevac})
and  POVM (\ref{povmi}) are both positive. Taking into account
the definition (\ref{rhoout}) of $\hat{\rho}_s$,
we immediately conclude that the Wigner function $W_s$ of the state
$\hat{\rho}_s$ must be positive.
In this context we can say that direct photodetection
by avalanche photodiode is, in certain sense, more non-classical
than balanced homodyning because the Wigner function of the operator
$\hat{\Pi}=\hat{I}-|0 \rangle\langle 0|$ is negative in some regions
of phase space.

\section{Imperfect homodyne detection}

So far, we have assumed an ideal homodyne detector with unit efficiency.
Let us now analyze the influence of
imperfect  detection with efficiency $\eta<1$
on the properties of conditional state of signal mode.
We can expect that the nonclassical properties of $\hat{\rho}_s$
will diminish with decreasing $\eta$ but we shall see that
the suppression of nonclassicality is not severe.

A realistic homodyne detector can be modelled by
an ideal homodyne detector whose signal input is preceded by a
beam splitter, where the signal mode is mixed with an auxiliary mode
which is in chaotic state. The detected quadrature can be thus expressed
as \cite{Leonhardt95}
\begin{equation}
\hat{x}_{i,\eta,\phi} =
\sqrt{\eta} \hat{x}_{i,\phi}+\sqrt{1-\eta} \hat{x}_{\rm aux}.
\label{xeta}
\end{equation}
In our case the quadrature $\hat{x}_{i,\eta,\phi}$ has again a Gaussian
distribution with zero mean and variance which does not depend on $\phi$,
\begin{equation}
\langle (\Delta\hat{x}_{i,\eta})^2\rangle=
\frac{1+2\bar{n}(1-\eta)+
\lambda[2\eta(1+\bar{n})-1-2\bar{n}]}{2(1-\lambda)},
\label{xvariance}
\end{equation}
where $\bar{n}$  is mean number of chaotic photons in
the auxiliary mode. Making use of the formula (\ref{xvariance})
 we can derive the probability that the absolute value
of the measured idler quadrature is higher than the threshold $x_0$,
\begin{equation}
C(\lambda,x_0,\eta,\bar{n})=1-{\rm erf}\left(\frac{x_0}
{\sqrt{2 \langle (\Delta\hat{x}_{i,\eta})^2\rangle}}\right).
\label{Cgeneral}
\end{equation}

In what follows we shall assume that the auxiliary mode is in vacuum
state, hence $\bar{n}=0$ and Eq. (\ref{Cgeneral}) simplifies to
\begin{equation}
C(\lambda,x_0,\eta)=
1-{\rm erf} \left(x_0\sqrt{\frac{1-\lambda}{1+(2\eta-1)\lambda}}\right).
\label{ceta}
\end{equation}
In the limit of ideal detector, $\eta=1$, this expression reduces to Eq.
(\ref{c}). We may utilize the formulas (\ref{nmom}),
(\ref{normalmom}) and calculate the moments of the
photon number distribution (\ref{pndistribution}).
After some algebra we obtain
\begin{eqnarray}
\langle \hat{n} \rangle &=&
\frac{\lambda}{1-\lambda}
+\frac{2\eta \lambda x_0 C^{-1}(\lambda,x_0,\eta)}{\sqrt{\pi}
[(1-\lambda)(1+(2\eta-1)\lambda)^3]^{1/2}}
\nonumber \\ & & \times
\exp\left(-x_0^2\frac{1-\lambda}{1+(2\eta-1)\lambda}\right),
\end{eqnarray}
\begin{eqnarray}
\langle :\! \hat{n}^2 \! :\rangle&=&
\frac{2 \lambda^2}{(1-\lambda)^2} +
\frac{2 \eta x_0 \lambda^2 C^{-1}(\lambda,x_0,\eta)}
{\sqrt{\pi}\left[(1-\lambda)[1+(2\eta-1)\lambda]^3\right]^{1/2}}
\nonumber \\[1.5mm] && \times
\left[\frac{4-3\eta+4(2\eta-1)\lambda}{(1-\lambda)(1+(2\eta-1)\lambda)}
+\frac{2 \eta x_0^2}{[1+(2\eta-1)\lambda]^2}\right]
\nonumber \\[1.5mm] && \times
\exp\left(-x_0^2\frac{1-\lambda}{1+(2\eta-1)\lambda}\right).
\end{eqnarray}

The POVM describing homodyne detection with realistic detector
can be expressed as
\begin{equation}
\hat{\Pi}_{i,\eta}=\int_{\cal{X}} \hat{\Pi}_i(x_\eta) d x_\eta,
\end{equation}
where $\hat{\Pi}_i(x_\eta)$ reads
\begin{eqnarray}
\hat{\Pi}_i(x_\eta)&=& \frac{1}{2\pi}\frac{1}{\sqrt{\pi(1-\eta)}}
\nonumber \\ && \times
\int\limits_0^{2\pi}
\int\limits_{-\infty}^\infty |x^\prime;\phi\rangle_i\langle x^\prime;\phi|
e^{-(x-\sqrt{\eta} x^\prime)^2/(1-\eta)}  d\phi d x^\prime.
\nonumber \\
\end{eqnarray}
The coefficients $q_n$ in the photon-number distribution
(\ref{pndistribution}) may be again
calculated in a similar way as in the case of ideal detector. After a
long but straightforward algebra we arrive at
recurrence relation for $q_n$,
\begin{eqnarray}
q_n&=&q_{n-1}+\frac{\eta}{\sqrt{\pi}} e^{-x_0^2}
\sum_{k=0}^{n-1} \left(\frac{\eta}{2}\right)^{n-k-1}
\frac{(1-\eta)^k}{(n-k)!}
\nonumber \\ && \times
{n-1 \choose k}  H_{n-1-k}(x_0) H_{n-k}(x_0).
\end{eqnarray}
where we start from $q_0=1-{\rm erf}(x_0)$.

The detrimental effect of $\eta<1$ on the suppression of photon-number
fluctuations can be deduced from Fig. 6.
In Fig. 6a we plot the probability $C(\lambda,x_0,\eta)$
of generation of the state with Poissonian fluctuations ($Q=0$).
We can see that the probability
reduces with decreasing $\eta$. Our numerical calculations reveal
that we cannot achieve $Q=0$ for any $\lambda$  when
$\eta<0.5$. Figure 6b shows the probability of preparation of a state
with sub-Poissonian statistics ($Q=-0.05$). As $\eta$ diminishes, the
maximum probability of generation decreases and the optimal $\lambda$
increases.

\begin{figure}[!t!]
\centerline{\psfig{figure=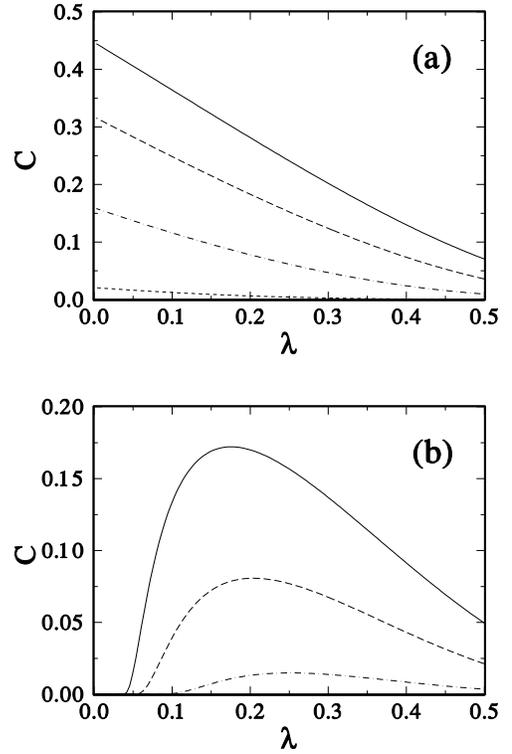,width=0.75\linewidth}}
\vspace*{3mm}
\caption{The probability of generation of state with Mandel $Q$-factor
(a) $Q=0$ and (b) $Q=-0.05$ is plotted for four different detection
efficiencies $\eta=0.9$ (solid line), $\eta=0.8$ (dashed line), $\eta=0.7$
(dot-dashed line) and $\eta=0.6$ (short-dashed line).}
\end{figure}

We can provide a simple argument explaining why nonclassical light
cannot be generated when $\eta<1/2$. To this point let us assume that
the homodyne detector in Fig. 1 is replaced with heterodyne detector
measuring projections to coherent states,
$\hat{\Pi}_i(\alpha)=\pi^{-1}|\alpha\rangle_i\langle\alpha|$. Suppose that
the output signal is accepted only when the absolute value of the real
part of measured $\alpha$ is larger than $\alpha_0\geq 0$.
Since the idler is in chaotic state, its $Q$-function is Gaussian and we
may express the probability that $|{\rm Re}(\alpha)|>\alpha_0$ as
\[
\tilde{C}(\lambda,\alpha_0)=\frac{2(1-\lambda)}{\pi}
\int_{-\infty}^\infty \int_{\alpha_0}^\infty
e^{-(1-\lambda)(\alpha_R^2+\alpha_I^2)} d\alpha_R d\alpha_I,
\]
\begin{equation}
\tilde{C}(\lambda,\alpha_0)=1-{\rm erf}(\alpha_0\sqrt{1-\lambda}).
\label{Ctilde}
\end{equation}
We can see that $\tilde{C}(\lambda,\alpha_0)\equiv
C(\lambda,x_0=\alpha_0,\eta=0.5)$ hence the photon-number statistics of
conditional signal state generated via heterodyne detection on idler
coincides with statistics of conditional signal state prepared by means of
homodyne detection on idler with efficiency $0.5$.
After the detection of $\hat{\Pi}_i(\alpha)$
the signal mode is projected to  coherent state with complex amplitude
$\alpha_s=\sqrt{\lambda}\alpha^\ast$ \cite{Watanabe88}.
Thus the conditional state prepared by
heterodyne detection on idler is a stochastic mixture of
coherent states and cannot exhibit any nonclassical features.
Assume now that the auxiliary mode in Eq. (\ref{xeta}) is in thermal state.
When we look at the formula (\ref{Cgeneral}), we find that
$C(\lambda,x_0,\eta,\bar{n})$ has the same form as
$\tilde{C}(\lambda,\alpha_0)$
in Eq. (\ref{Ctilde}) when
\begin{equation}
\eta=\eta_{\rm th} \equiv \frac{1+2\bar{n}}{2+2\bar{n}}.
\end{equation}
If $\eta<\eta_{\rm th}$, then the balanced homodyning is formally
equivalent to heterodyning with detection efficiency lower than unity.
Thus we conclude that the sub-Poissionian light can be
conditionally generated only when $\eta>\eta_{\rm th}$
holds. The threshold efficiency $\eta_{\rm th}$ is lowest when the auxiliary
mode is in vacuum state and $\eta_{th}=0.5$. The threshold $\eta_{\rm th}$
increases with $\bar{n}$ and approaches $1$ for large $\bar{n}$.

In the experiments, the efficiency is  reduced mainly due to only
partial overlap between the measured mode  and  local-oscillator mode.
If the overlap is good, then $\eta$ can be very high. For example,
in the single-mode  optical homodyne tomography \cite{Schiller96},
the efficiency $82\% \pm 5\%$ was achieved,
which would be fully sufficient for our purposes.
However, we deal with two-mode system and we need to extract the signal
mode which corresponds to detected idler. In the recent experimental
realization of the two-mode homodyne tomography \cite{Vasilyev00},
the estimated efficiency was only $35 \%$, which is below our threshold.
Nevertheless,
even such data may be used to demonstrate the suppression of noise via
conditional generation, although the photon number fluctuations could not be
reduced below the Poissonian level.

\section{Conclusions}

We have proposed a simple scheme for conditional generation of
sub-Poissonian light from the two-mode squeezed vacuum.
The scheme involves homodyne detection on the idler mode
and the experimental run is accepted only if the absolute value of the
detected quadrature is higher than certain threshold.
As the threshold is increased, the photon-number fluctuations of
conditionally generated state of signal mode are
gradually suppressed but the probability of generation decreases.
If we wish to prepare a nonclassical state with a given negative Mandel
$Q$-factor, then we can choose an optimal squeezing parameter $\lambda$
leading to maximum generation probability.
Nonclassical light can be prepared only when the efficiency
of  homodyne detection $\eta$ is higher than $\eta_{\rm th}\geq 1/2$.
Although the threshold $1/2$  was not achieved in the
recent experiment on two-mode homodyne tomography \cite{Vasilyev00},
the condition $\eta>\eta_{\rm th}$ is by no means severe and
the present method may become experimentally feasible in a near future.

\acknowledgments
I would like to thank J. Pe\v{r}ina, R. Filip and L. Mi\v{s}ta, Jr.
for valuable comments.
This work was supported by Grant LN00A015 of the Czech Ministry of
Education.

\end{document}